# Application of LEAN Principles to Improve Business Processes: a Case Study in Latvian IT Company

Anastasija NIKIFOROVA[1], Zane BICEVSKA[2]

Faculty of Computing, University of Latvia, Raina bulv. 19, Riga, LV-1586, Latvia

[1]ORCID: 0000-0002-0532-3488, [2] ORCID: 0000-0002-5252-7336

`an11093@lu.lv, Zane.Bicevska@di.lv`

**Abstract**. The research deals with application of the LEAN principles to business processes of a typical IT company. The paper discusses LEAN principles amplifying advantages and shortcomings of their application. The authors suggest use of the LEAN principles as a tool to identify improvement potential for IT company's business processes and work-flow efficiency. During a case study the implementation of LEAN principles has been exemplified in business processes of a particular Latvian IT company. The obtained results and conclusions can be used for meaningful and successful application of LEAN principles and methods in projects of other IT companies.

## Introduction

Nowadays implementation of sophisticated IT projects have become necessity in many industries as well as in public sector; however, relatively small part of IT projects prove to be explicitly successful and cost effective from an IT company's perspective (Ko et al., 2005). Different approaches and methods are exploited by IT companies in an effort to improve business performance and cost effectiveness while complying with high product or service quality standards and without compromising relevant deadlines. LEAN thinking and management is one of such approaches. Many American and European companies have testified improvement of their performance due to the introduction of LEAN principles. There are also numerous case studies on application of LEAN principles in different industries and sectors: automotive industry (Soderquist and Motwani, 2010), (Vinodh et al., 2011), healthcare (Radnor et al., 2012), (Toussaint and Berry, 2013), textile industry (Hodge et al., 2010), call centres (Laureani et al., 2010), etc.

Though there is only limited information available regarding usage of the LEAN approach in service industries (Apte and Goh, 2004), and a very few information can be found about application of LEAN principles in the Baltic countries (Miina, 2013), (Mahmood, 2015). The IT industry is mostly analysed by using examples of Indian software companies (Staats and Upton, 2009), (Staats et al., 2011), (Deep et al., 2017).

According to the LEAN approach, the ultimate goal of every business should be customer satisfaction with the service or product. Therefore, a lean manufacturing company should perform only those activities which add value to service or product, comply with customer requirements while meeting high quality standards and minimizing waste and any kind of redundancy. Many principles and methods such as Kanban, 5S, 5 why's, value stream mapping, Kaizen and others have been developed under LEAN premise.

Due to the experience shared by the foreign companies the authors consider introduction of LEAN approach to be a meaningful attempt in order to find a solution for a series of performance issues faced by the Latvian IT companies. The authors offer thorough exploration of LEAN approach amplifying advantages and shortcomings of LEAN implementation in a particular IT company. The authors also elicit effects the LEAN approach might have on delivered IT product and service quality in the areas such as computer network administration, hardware maintenance, user administration, IT guarantee and support service, etc.

Within the framework of this research the authors intended creating of value stream mapping for specific IT projects. Business process models and value stream mapping (VSM) methods were used to capture IT company's workflow activities thus creating foundation for application of LEAN principles. Workflow activities were analysed applying LEAN principles in order to identify activities that create no value for the product or service. Results were used to redefine requirements for workflow activities to eliminate waste and enhance improvements.

The sections of this paper include: an overview of LEAN approach, its pros and cons, derived principles and applicability (Section 2), a short analysis of business processes of a particular Latvian IT company (Section 3), analysis of LEAN wastes and their detection in the Company (Section 4), evaluation of LEAN principles contribution to company's business processes and specific IT projects (Section 5).

## 1. Project success rate

Ever since the very beginning of the IT industry, the desirable IT project outcome at a reasonable cost has been an issue of relevance. A significant amount of statistical data proves the same fact: the large part of IT projects experience rather low success rate. The statistics also show that this trend has been persistent in the course of time.

The Standish group has been collecting information on project success rate for more than 20 years. In 2015, they have published Chaos Report (The Standish Group, 2013) where projects were divided by their success rate into three groups: successful, failed and challenged. The report defines projects which are finished within time and budget, and which result satisfy customer as successful projects. Challenged projects are those were the goals are achieved and tasks are completed and approved but they are late, over budget, and have fewer features and functions than originally specified. Failed projects are interrupted projects, i.e. they weren't completed.

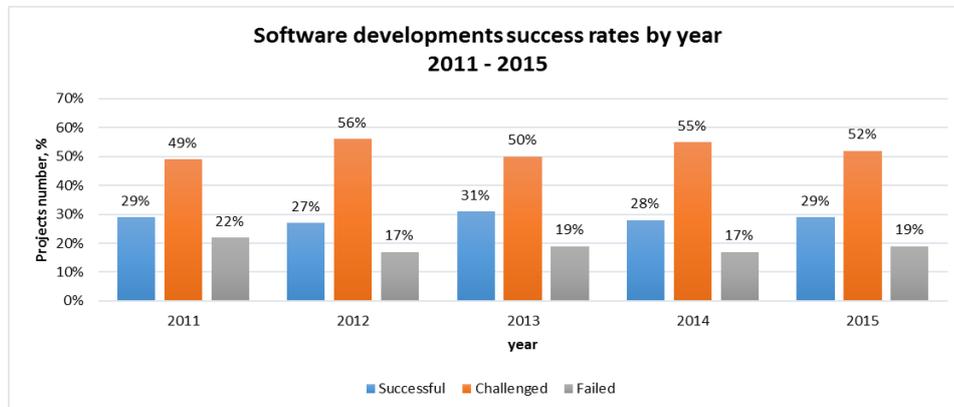

**Fig. 1.** Software development success rates by year

As illustrated in Figure 1, which is summary of (Hastie and Wojewoda, 2015), during the time period from 2011 to 2015 the number of successful projects was relatively small and varied from 29% to 30% of all software development projects. Within the same period on average 52% of software projects were challenged and around 19% had failed. It is remarkable that the breakdown among project success groups fluctuates rather insignificantly within the given period. This statistical data show that the aim of finding a trend braking solution that would improve overall software development project success has not been accomplished yet.

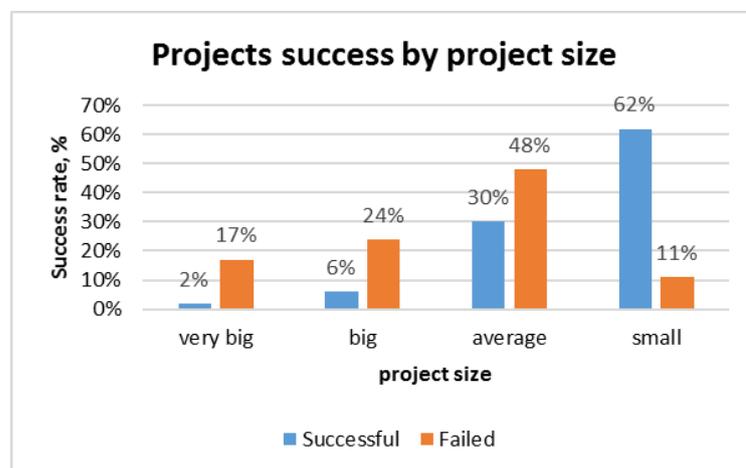

**Fig. 2.** Project success by project size

Figure 2 (created according to (Hastie and Wojewoda, 2015) shows that size of a project is significant factor to project success. Moreover, the smaller the project is, the higher is probability that project will be successful. Projects with less than 5 team members and less than 3 months of duration show only 11% failure rate. However, large and very large projects with teams of more than 30 people and project duration of more than 1 year produce on average only 4% success. The overwhelming majority of the large projects were challenged (Hastie and Wojewoda, 2015).

Project Management Statistics 2015 (Bonnie, 2015) show that on average large IT projects exceed the planned budget by 45%, overrun the scheduled time by 7%, and deliver 56% less value than expected (Project Management Institute, 2015), Furthermore, 17% of large IT projects (with budget exceeding $15M) go so badly that they even threaten business sustainability of the company (Bloch et al., 2012).

Moreover, almost 20% of all IT projects regardless of their size have cost overrun of an average 200% and schedule overrun of 70% (Flyvberg and Budzier, 2011).

There is a long list of reasons among them unclear project objectives, lack of business focus, unclear software requirements, resource trashing, unrealistic schedule planning, changing requirements, underestimated technical complexity, personnel issues, missing skills and competencies, conflicting priorities and others that might cause an IT project to fail (Project Management Institute, 2015), (Bell and Orzen, 2011).

Many IT companies try to improve their project success rates by adopting project management approaches such as Agile or Lean, or applying development methodologies such as Scrum thus expecting to achieve better project results and success. When comparing IT projects carried out according to a certain project management methodology and the projects that do not follow any standard methodology (Bonnie, 2015), the statistics show that 38% of the projects with methodology deliver project outcome within the planned budget and 28% of them within the desired timeframe; while only 31% of the projects without management or development methodology meet budget requirements and 21% stay on schedule. Moreover, according to (PricewaterhouseCoopers, 2012) where largest participating industry was Information Technology, the quality of the first group's deliveries is higher by 8% (PricewaterhouseCoopers, 2012).

Many management philosophies and development approaches promise to make IT projects successful. The following breakdown shows percentage of IT projects that use project management approach and/or development methodology or their combination: 41% use PMBOK, 9% use an IT methodology, 8% use a combination of methods, 9% use another approach, 4% use an in-house method, 3% use PRINCE2 and 26% don't use any methodology (PricewaterhouseCoopers, 2012).

The most popular project management and development approaches used by IT companies are PMBOK (Project Management Institute, 2017), Agile, plan-driven approaches such as waterfall, V-model, linear, b-model, evolutionary and incremental software developments (Pressman and Maxim, 2015).

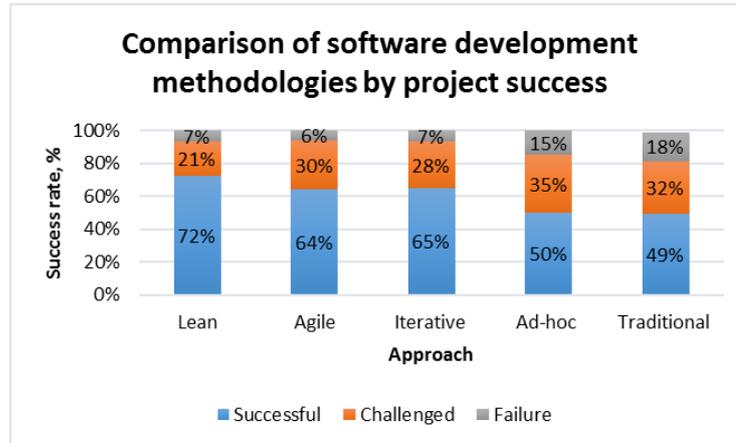

**Fig. 3.** Comparison of software development approaches by project success

According to Figure 3 (created according to (WEB, e)), IT companies that have adopted LEAN approach have accomplished more successful projects than companies which follow other approaches. Analysis of relation between project success rate and development methodology suggests that projects carried out under LEAN approach have the highest success rate of 72%, while only 21% of LEAN projects are challenged and merely 7% have failed.  The next best result provides AGILE approach with 64% successful projects, 30% challenged projects and merely 6% failed projects. Projects carried out according to iterative, ad-hoc and traditional approaches show 7-23% lower project success rate than LEAN projects. Unfortunately, there is lack of statistics on project success rates in Latvian IT companies especially in relation to applied project management and development approach. However according to publicly available information Latvian IT companies mostly use Agile approach.

In further section, this paper will explain LEAN philosophy, principles, advantages and possible improvements LEAN can bring about in an IT project implementation and an IT company in general. The authors will exemplify application of LEAN principles in a successful Latvian IT company (further the Company) with an aim to improve its business processes.

## 2. LEAN

Lean is a concept of thinking an acting to organize human activities to deliver more benefits to society and value to individuals while eliminating waste. LEAN and especially LEAN IT is more than a practical set of tools, techniques, and practices. LEAN thinking principles determine organization's philosophy thus enabling behavioural and cultural transformation throughout the organisation. LEAN also looks at the role of quality in the value creation for the customer from different perspective developers are focusing only on those activities which will add value to developed product from the customers perspective, ignoring all activities for which customer wouldn't be ready to pay.

Application of LEAN principles helps the organization to achieve operational excellence and provides tools to enable continuous improvement of IT operations, value creation and high-quality IT services and products (Bell and Orzen, 2011). The LEAN thinking is being adopted for the purposes of value creation, and to examine the potential role of the customer in improving supply chain performance.

LEAN foundation consists of three components: constancy of purpose, respect for the individual and pursuit of perfection. Such attributes as proactive behaviour, customer expectations, systemic thinking and quality at the very source and throughout the workflow are vital in LEAN vision to enhance flow/pull/JIT improvement while culture represents the capstone of LEAN (Bell and Orzen, 2011).

## 2.1. LEAN principles

The LEAN approach recommends implementation of 5 key principles in every organisation or project:
1. Identify value: identify what creates value for customer in a supply chain or development project;
2. Map the value stream: create a visualisation of activities and processes;
3. Identify wastes: identify all possible wastes which could affect work efficiency and provided results;
4. Eliminate wastes: eliminate all in previous step identified wastes by applying one of the LEAN tools or techniques, prevent their reoccurrence.
5. Pursue perfection: continuously improve results even in case of success because this LEAN principle facilitates striving for perfection.

**Table 1.** LEAN software development. Principles and tools

| No. | Principle | Tools |
|---|---|---|
| 1. | Eliminate waste | Seeing waste (will be discussed in Section 4.1), value stream mapping (see Section 4.2). |
| 2. | Amplify learning | Feedback, iterations, synchronization, set-based development |
| 3. | Decide as late as possible – defer commitment | Options thinking, the last responsible moment, making decisions |
| 4. | Deliver as fast as possible | Pull systems |
| 5. | Empower the team | Self-determination, motivation, leadership, expertise |
| 6. | Build integrity in | Perceived integrity, conceptual integrity, refactoring, testing |
| 7. | See the whole – optimize the whole | Measurements, contract |
| 8. | Instructions and warranty | Recommendations: instructions, troubleshooting guide and warranty |

With the growing popularity of LEAN approach in many industry sectors, including IT, Poppendicks (Poppendieck and Poppendieck, 2003) formulated 8 principles of LEAN software development supplementing them with specific tools to be used by the business. Authors of this paper examined these principles in detail and suggest that the most part of these principles can be usefully applied also to other kinds of IT projects providing little adaptation to each specific kind of IT project, for instance, the principles of software testing and verification can be easily adapted to IT hardware or IT service (such as computer network administration) delivery when hardware components or user permissions are tested and verified whether or not they comply with a requirement, specification, etc. These principles and LEAN tools are summarised in the Table 1.

However, some of these principles may raise discussion, for instance, the principle of "decide as late as possible" might be considered as too risky since the dividing line between "late as possible" and "delayed" decision is rather arbitrary. Inappropriate application of this principle might cause new problems and even more (time) wastes. Though a proper application of this particular principle can be useful to minimize risks and avoid refactoring.

Almost all of the aforementioned 8 principles, often called "success factors", agree with 9 principles discussed in (van der Zee et al., 2015). They are adapted to IT business specifics, and the principles are divided into three groups - (1) strategic, (2) operational and (3) tactical:

1. Strategic: constancy of purpose, pursuit of perfection which along with the respect for the individual comprise the very fundamentals of LEAN. In an organization that applies these strategic principles treats all employees, customers and their opinions with respect, captures and exercises their potential, thus contributing to their develop and enhancing continuous improvement.
2. Operational: proactive behaviour, taking responsibility for own and colleagues activities.
3. Tactical: quality guarantee and maintenance, built-in quality, customers opinions establishment to define goals and requirements according to his vision, flow creating to guarantee continuously process flow eliminating any delays or waiting and systems thinking guarantee (Bell and Orzen, 2011).

In (Bell and Orzen, 2011) authors describe that mere appliance of specific tools and techniques is less efficient than implementation of LEAN principles in an organization and its business processes, especially principles which are relevant to organisational culture and promote respect for colleagues and customers thus fostering long-term results. Tools and techniques are most effective when used only after these principles are implemented and well established in an organisation.

## 2.2. Comparison with AGILE

LEAN approach is similar to the AGILE methodology which is widely used in IT industry. AGILE is the subset of LEAN and both being the subset of systemic thinking (Robinson, 2013), (Nedre, 2018) which share in common the following principles:

1. respect for colleagues and customers; continuous improvement; iterative development which is in core of AGILE aligns with the Lean principles of "deliver as fast as possible" and "defer commitment";
2. short feedback loops between AGILE developers and their stakeholders help teams create a habit of eliminating processes, activities, and products that do not directly result in customer value which corresponds with the LEAN principle of "eliminate waste";

      3. AGILE disciplined project management process is similar with the Lean "Build quality in" principle according to which any tedious, repeatable process or any process that is prone to human error must be automated and standardized which allow LEAN teams to error-proof significant portions of their processes, so they can focus their energy on creating value for their customers.

Relying on a consistent, disciplined process allows Agile teams to continuously refine and optimize their processes for value delivery (WEB, b).

Moreover, SCRUM developers admit that LEAN inspired them to develop SCRUM thus explaining the obvious similarity of principles.

However, there are also some differences between the AGILE and LEAN approaches, for example,
      a) While AGILE follows the "doing" principle aiming at fast deliveries and adaptation to changes, LEAN's aim is to work "smarter" by improving process, reducing everything that doesn't add value to customer thus following the "thinking and doing" principle.
      b) The development process according to AGILE is flexible emphasizing reaction to changes according to "do-inspect-adapt" principle but LEAN development process is sustainable emphasizing planning and responding according to "inspect-plan-do" principle.
      c) LEAN focuses on process while AGILE focusses on people.
      d) LEAN projects are realized by a working unit - team but AGILE projects by interacting individuals.
      e) According to LEAN, rework in design adds value, but in making it is waste. In AGILE any kind of up-front work must be minimized and code must be reworked to get quality.
      f) While AGILE considers face-to-face communication to be the most effective method of conveying information within development team, LEAN prefers visual management as the main communication method, however face-to-face communication is very important, too.

It shows that these two approaches are similar however not identical, moreover, there are certain differences in basic principles and philosophies that lead to different results.

## 3. Analysis of the Company

The LEAN principles were described and exemplified by a Latvian IT Company which aims to improve work efficiency, rate of project success and the effectiveness of business processes. The Company's main activities are in the IT service areas such as custom software development, network administration, hardware maintenance and sales. Therefore introduction of LEAN approach in the whole company can be regarded as challenged due to different kinds of IT services.

The project success rate achieved by the Company is extremely high – approximately 97% in comparison with average rates in industry (Hastie and Wojewoda, 2015), (WEB, d). Only in very rare cases projects have failed or have been challenged (the number of challenged projects is less than 3-5%). Nevertheless, the goal of this research is to prove that introducing LEAN approach in the Company there is a potential to improve even high project success rates effectiveness of business processes and overall work efficiency.

At first a thorough analysis of the Company's was carried out in order to detect whether there are any kind of process or project settings that already work according to LEAN. As a result of analysis authors have identified a number of LEAN features such as:

- focus on customer;
- manufacturing according to the customers demand;
- intense communication with customer at project initialization setting goals from customer's point of view and during development in order to identify requirements, needs and preferences;
- sometimes employee might not have tasks – may be free because workload depends on customers demand and even task assignment is realized according to employee workload;
- employees are multi-skilled, i.e. besides their educational degree in computer science they have competencies with approved certification, they also carry out tasks which might exceed their competency requirements thus showing their potential and capturing opportunity to grow professionally;
- testing is a significant part of development activities and takes part also prior to product delivery thus enabling early detection of problems and ensuring high quality delivery to customer. Moreover, prevention of repeated problems is significant part of improvement process within framework of organization's quality management system;
- there is a minimum number of started and not ended tasks – "work in progress" (WIP). Usually tasks are carried out consecutively. In cases when several customers would require instant reaction to issue notification or an incident affecting several customers would occur then different problem solving tasks are carried out simultaneously. In some cases, problem solving process requires customer involvement providing necessary information. However, according to the most Service level agreement (SLA) terms problem solving deadline is 30 days (Maskell et al., 2011), (Nagar, 2017);
- permanent improvement (product quality, processes, work efficiency) process is in place as a component of the quality management system;
- the Company has standardized work processes, i.e., all relevant processes are designed within quality management system and are available in graphical form;
- result of preceding task, or necessary spare part or resource is delivered exactly when it is necessary, i.e. waiting time is minimized – observance of JIT principle;
- task and issue tracking is effectively used, i.e. tasks have completion deadlines – observance of "takt time".

It is important to identify the existence of LEAN features thus helping to select and apply LEAN tools and techniques in the Company's business areas where they are most useful and bring the most results.

Since almost 15 years the Company has quality management system certified according to the *ISO 9001:2015*. The existence of this well-established quality management system is one of the main reasons of high performance results. The set quality goals include customer satisfaction meeting requirements, strong customer focus, the employees' motivation and strong commitment by the top management. The process approach and continual improvement which are the key principles of the quality management system are very similar to the LEAN and thus prove the existence of LEAN features in the Company (WEB, a). The existence of this standard or its elements in an organization contribute to the excellence models (one of these models is LEAN) implementation (Leilands, 2009).

As far as many Lean features are used in the Company, authors considered that there are necessary only few strategic and tactic changes such as improvement principle which includes identifying and reducing of wastes (is used just partially), the flow principle and systems thinking which includes value stream development and human respect.

The next chapter deals with relevant processes in the Company trying to identify wastes or non-value-added activities as well as potentials to reduce such wastes and activities.

## 4. LEAN wastes

According to the LEAN, there are 3 problems which can be observed in each enterprise: 'muda' or wastes which are explained in this section, 'mura' (un-evenness) and 'muri' (overburden) which were firstly defined by Toyota (WEB, b).

To understand these terms better, let us define each of them: 'muda' or waste is any activity that consumes resources but creates no value for a customer; 'mura' or un-evenness is a variation in the operation of a process not caused by the end consumer, 'muri' is overburden on equipment, facilities and people, caused by 'mura' and 'muda' (Womack, 2008).

It is extremely important to identify and to reduce activities that do not add value or do create waste. A non-value adding activities and wastes are not necessary to the outcome developed according to the customer's requirements, and customers are not ready to pay for these activities.

There are 7 wastes of LEAN also called *muda*: overproduction, transportation, waiting, motion, inventory, extra processing and defects. Wasted creativity or potential of employees is an additional kind of wastes which is not less important, however usually is not included in the list of wastes.

While all these "7+1" wastes are common for the most organizations in all industries, Poppendiecks in (Poppendieck, 2003) formulated IT industry - specific wastes and described then by changing names of some previously mentioned wastes in order to make them more understandable. For example, overproduction in software development means development of additional functionality or extra features, transporting includes not only movement but also switching between tasks, partially done work is considered inventory. Table 2 compares description of wastes according to LEAN and description of wastes in IT industry-specific terminology.

**Table 2.** LEAN wastes

| Waste | Description | Waste in IT |
|---|---|---|
| inventory | storing parts, documentation, equipment which maybe someday will be necessary – ahead of requirements | inventory, partially done work |
| extra processing; over processing | rework, excess inventory, reprocessing, etc. | extra activities, extra processes |
| overproduction | producing more than customer needs according to forecasts (producing before it is needed) – making more than is immediately necessary | extra features |
| transportation | movement of tools, products, people and information | task switching, transportation |
| waiting | "work-in-process" when no value is added; waiting for parts, response, information, delivery | waiting |
| motion | unnecessary movements of people - turning, reaching, lifting | motion |
| defects | rework, incorrect documentation | defects and their reduction |

The authors of (Poppendieck and Poppendieck, 2003) have described the aforementioned wastes and have developed the following recommendations for removing of wastes:

1. Inventory and work components: in case when a task or a final product contains several components developed separately there is a certain risk related to compatibility of components when integrated into one product even if each component has passed through testing successfully.
   Possible decision: minimize work done in components to reduce risks and wastes. In this case LEAN provides supportive tools and techniques, for example, "just-in-time" (JIT), continuous flow, takt time, SMED and Kanban discussed in more detail in Section 5.
2. Extra processing activities: extra processing activities include paper work, development of documentation. According to LEAN, the paper processing can cause waste of time and resources especially in cases when purpose or usage of documentation is not clear or value created by documentation does not justify the spent resources. LEAN defines paperwork as a time and resources consuming activity, that might cause latent quality problems, slow down response time and become obsolete. Therefore, to avoid extra processing activities it is important to evaluate whether documentation required by customer will give an additional value and provide the necessary support in usage, decision making or communication, or both. The key issue is to clarify whether customer and developer would not benefit more by refusing to do extra processing actives.
   Possible decision: identify and eliminate unnecessary documentation development. In cases when customer's requirement cause development of

    extra documentation, try to clarify purpose of documentation and explain possible wastes. In cases when the documentation is considered necessary, for example, Software Requirements Specification, Software Design Document, User Manual, Requirements traceability matrix, Use cases, etc., LEAN provides three rules how to minimize risks and wastes, and add value to documentation: (1) keep it short, (2) keep it high-level design, (3) do it off-line. Thus, documentation that would qualify as value-adding activity would meet the following criteria: a well-formatted and compact document providing only the necessary information, user friendly, enabling fast updating, searching and checking for completeness. LEAN also recommends using Kaizen to reduce other wastes related to documentation.

3. Extra features: all features which aren't required by customer should be considered as the extra features which almost always qualify as wastes. Development of extra features or provision of extra service requires time and human resources throughout the whole product or service lifecycle. Moreover, there is a high probability that the extra added features will become obsolete. According to CHAOS researches (The Standish Group, 2013) approximately 50% of all features required by customers often are not utilized by users.
Possible decision: LEAN advises to deliver only those features which are necessary to customer according to his requirements and purpose, avoiding any extra features that do not have clearly defined usage.
Traditionally, LEAN set the following three rules (1) produce exactly what customer needs, (2) produce only as many as it is necessary for customer (3) produce in time when customer needs it. In order to fulfil these rules LEAN tools and techniques such as "takt time", Kanban and SMED are recommended.

4. Task switching: work-in-process occurs assigning multiple projects or multiple tasks within one project to employees. From LEAN perspective waste of time is created when employees switch between tasks (projects) including switching between software, task context etc. Simultaneous multitasking creates too many interruptions, therefore LEAN suggests consecutive task performance.
Possible decision: avoid assigning multiple tasks covering different projects to employee. It is also advised to use Kanban, day and week boards and follow "deliver as fast as possible" principle, continuous flow, theory of constraints and value stream mapping.

5. Waiting: different kinds of delays such as waiting for customer response, approval or evaluation, testing results, component delivery, employee availability, project start-up, documentation or any kind of uncertainty (e.g. legislation) result in delayed project outcome or service delivery.
Possible decision: in cases of uncertainty LEAN suggests several options, for example, "decide as late as possible" principle, continuous flow and standardized work.

6. Motion: rendering of services in IT companies are usually organized in project teams. Communication among team members is crucial for better understanding of context and tasks, as well as planning, risk evaluation and defining prevention measures, i.e., every project aspect to ensure successful project outcome, therefore, moving around in order to discuss task or project

      related issues with team members located in different sites may cause huge time wastes. It also refers to hardware maintenance service when motions are necessary to find an item or spare part to perform activity at workplace.

      Possible decision: co-locating all team members at the same physical location enables face-to-face communication thus reducing time wastes. LEAN suggests organization of employee's workplace according to 5S LEAN technique and value stream mapping in order to detect motion waste, necessary or unnecessary accessories or equipment.

7. Defects and their reduction: defects and errors may occur in production, especially in software and hardware production. Releases of new versions, upgrades and updates are common means to reduce errors and defects. Flaws of defects detected in early design and development stages can easily be eliminated without significant financial loss, however defects identified during later stages of product life cycle, e.g., after delivery might cause significant financial loss and even damage reputation. Therefore, it is important and sometimes even crucial to identify defects during early stages and reduce them as fast as possible but in the best case take precautions to avoid flaws.

   Possible decision: immediate testing of completed task, frequent integration in order to check compatibility of components, and product release as soon as possible.

It is important to mention that other approaches do not consider the aforementioned wastes as such, for example, Kano model which is widely used in many approaches such as Agile and promises to achieve the best result and customer satisfaction says that additional functionality is necessary in order to guarantee customer satisfaction (even delight), to attract new customers, to deliver unique product. According to the Kano model there are three types of requirements: must-be, one-dimensional and attractive requirements. "Attractive requirements" are neither explicitly expressed nor expected by the customer. Fulfilling these requirements leads to more than proportional satisfaction (Sauerwein et al., 1996). However, according to LEAN here lays one of the most dangerous wastes since the development of additional, i.e., unnecessary functionality requires also additional resources such as time, work effort, finances, etc. (WEB, f).

While according to LEAN development of documentation is one of the non-value adding activities, other approaches and methodologies claim to have proved that projects success often is based on qualitatively developed documentation. Moreover, documentation is used to provide customer with information on work/project progress and success. Therefore, is important to understand that LEAN does not deny documentation as such, but recommends development of documentation only if it provides value to customer.

The results of thorough analysis of the Company's processes and wastes are presented in the Table 3. Wastes in the Company's processes were divided into 6 groups and evaluated according to the following scale: 0 – waste was not detected, 1 – waste or its features were observed, however its impact was not observed, 2 – waste was observed, but its impact was minimal, 3 – waste was observed, but its impact was irrelevant, it should be reduced, 4 – waste was observed, it impacts work efficiency, it must be reduced, 5 – waste was observed, its impact on work efficiency is critical, waste must be reduced immediately before it will effect project success.

**Table 3.** Evaluation of the Company's wastes

| Waste | Validity | Degree of impact |
|---|---|---|
| extra features | ✗ | 0 |
| defects and their reduction | ✓/✗ | 1 |
| waiting | ✓ | **3** |
| extra processes | ✗ | 0 |
| inventory and partially done work | ✗ | 0 |
| task switching and transportation | ✓ | **4** |
| motion | ✓ | **4** |
| creativity or wasted potential of people | ✓/✗ | 1 |

The authors developed recommendations for each waste which had received more than 0-point. The aim of these recommendations is to reduce waste of the particular work activity/process or at least minimize its impact on the Company's processes and work efficiency:

1. Task switching: better daily and weekly task planning and assigning to team members considering their workload and using:
   - day-board or Kanban with one-day work plan for every employee. Plan should include column to record problems and incidents which appeared while carrying out specific task. 15 minutes plan discussion team meeting.
   - week-board with the list of projects in progress, their goals, tasks and responsible employees. Week plan is discussed every Monday and Friday to compare plans and results and solve the issues recorded on Kaizen board.

Usage of these boards enable faster decision making, gaining overview of team's workload thus increasing efficiency of project work.

2. Motion: co-locating all team members at the same physical location enables face-to-face communication thus enhancing collaboration such as discussing work results, getting comments, explaining and understanding tasks, getting necessary information. Face-to-face communication minimizes risk of false assumptions, task misunderstanding, errors, and defects.
3. Waiting: finding out the most appropriate way of communication or the most effective way of getting the necessary information thus reducing the waiting time. Minimize waiting for response and necessity of unplanned communication while clarifying customer requirements, task details in order to ensure "delivery fast as possible".
4. Defects and their reduction: making testing integral part of development, detecting and repairing faults immediately.
5. Creativity or wasted potential of people: creating employees' skills and knowledge matrix where each employee is evaluated by two factors: (1) readiness to help and to train/instruct colleagues and (2) skills.

   Such a matrix provides overview of available skills and competencies and which additional skills and knowledges are necessary and, moreover, whether the needed skills, knowledge and competencies can be derived from inside the Company, i.e., form other employees.

## 5. Applied LEAN tools and techniques

The authors suggest the usage of LEAN tools which were formulated by Womack's (Womack, 2008) in order to improve software development process also in other IT-specific projects. Considering the wastes which were explained in Section 3, the applicable LEAN tools are: value stream mapping, feedback, iterations, synchronization, set-based development, options thinking, the last responsible moment, making decisions, pull systems, queuing theory, cost of delay, self-determination, motivation, leadership, expertise, perceived integrity, conceptual integrity, refactoring and testing.

As it was already mentioned, the first LEAN principle is value stream mapping. This principle allows identifying wastes and non-value adding activities throughout the whole life cycle from acquisition to final delivery and maintenance. It provides visualisation of the whole process – the sequence of process steps, activities including deadlines and waiting between process steps.

According to LEAN visualisation is the best way to provide understanding and overview of situation, as well as presenting and explaining the issue to be solved.

**Table 4.** LEAN tools and techniques

| Tool, technique | Description | Focus |
|---|---|---|
| 5S | Steps of a workplace organization process, which maximise efficiency, reducing time and necessity spending time looking for an item or a tool which is necessary for specific activity, cleaning and organising it, which is based on 5 "s": sort (jap. seiri), set in order or straighten (jap. seiton), shine (jap. seiso), standardize (jap. seiketsu), sustain (jap. shitsuke) | waiting, defects, transportation, motion |
| 5 why's | Root cause analysis by repeatedly asking "why" going in depth of problem to resolve it and prevent its reoccurrence | defects |
| Pareto | Problem solving and identifying which problem harms business the most, setting focus on business improvement | all |
| Kanban | Implementing a pull production to reduce cost, inventory reduction and increasing ability to adapt to market changes | inventory, waiting, overproduction, defect |
| Kaizen | Continuous improvement process where all employees are involved working proactively and eliminating wastes | all |
| Value stream mapping | Visual mapping of workflow, identifying potential to reduce lead times and unnecessary process steps thus highlighting the opportunities for improvements | all |
| Day board | Supporting the start of the day's activities by 15 minutes meetings to discuss operational performance, goals and tasks and synchronize workload among team members. Each employee records all the activities he is going to do at a specific day. | all |
| Week board | Providing information on the team's objectives for the week, retrospective of the previous week achievements, possible improvements which are recorded on the Improvement Board; issues occurred, causes and possible solutions, necessary decisions | all |
| TPM (Total Productivity Maintenance) | Focuses on the machines used in production regularly improving their efficiency | waiting, defects |
| SMED (Single Minute Exchange of Dies) | Reducing the changeover time to less than 10 minutes that enables manufacturing in smaller amounts, reduces inventory and improves customer responsiveness | inventory, waiting, defect |
| PDCA (plan-do-check-act) | Method of proposing a change in a process, implementing the change, checking and measuring the results and taking appropriate action | all |
| Jidoka | Stopping a process automatically when a problem occurs thus reducing inventory and space due to pull instead of push production | waiting, inventory, overproduction |
| JIT | Implementing pull production based on customer's demand not pushing it on projected demand | waiting, inventory, overproduction |
| Gemba | Problem identification and solving by going back to the value creation "site" and problem has occurred | defects |
| Poka-Yoke | Eliminating defects by error proofing processes, making it impossible to make mistake | defects |

Fig. 4 shows value stream mapping for an average project realized by the Company including the following steps: registered customer request in issue tracking tool and management system, revision of the request by the responsible employee (project manager), acceptance or decline of the request. If it is accepted, the request becomes a task assigned to project manager. The project manager might revise the task, divide it into subtasks, add relevant details to task description, set deadline and assign the task (subtasks) to the responsible team member(s). The completed task is tested by different team member. If testing was successful, next step is to compilation of delivery by the project manager following the delivery to customer. In average, this process takes 1045 minutes, i.e., approximately 2 days. The value-adding activities take-up 920 minutes of these 2 days but waiting time is 135 minutes (see Fig. 4).

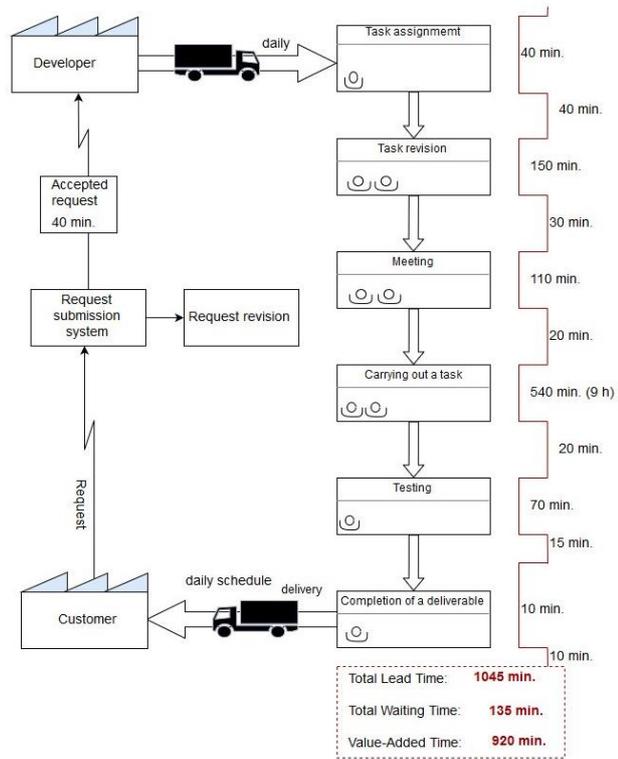

**Fig. 4.** Value Stream Mapping – an average project

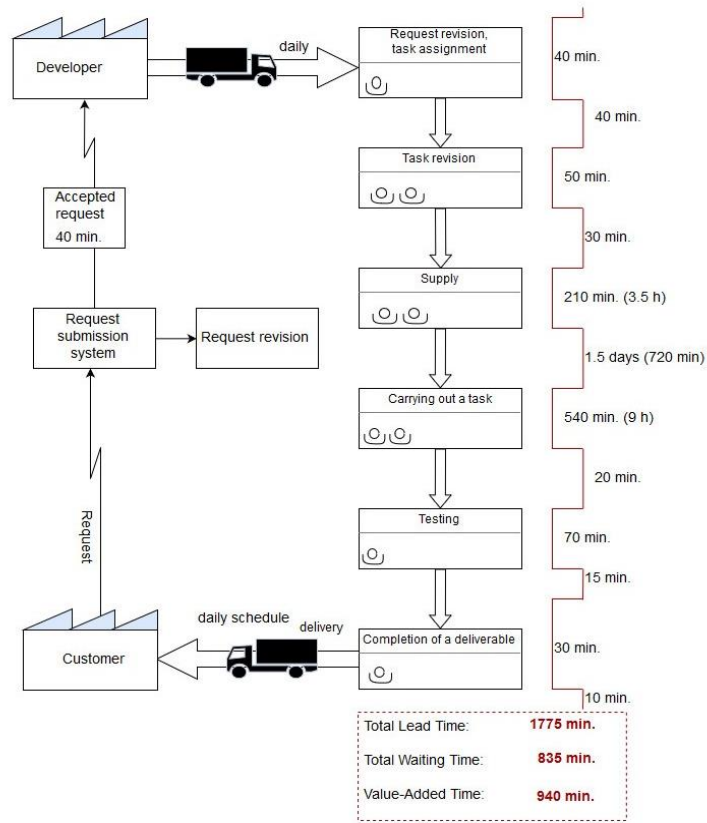

**Fig. 5.** Value Stream Mapping. Hardware maintenance and marketing

In case of hardware maintenance and marketing total task processing time as well as value-adding and waiting time slightly differed (see Fig. 5), therefore, it was necessary to draw a separate value stream map for every type of task.

According to the goal set by the authors the application of the value stream mapping enabled to identify potential for reducing the total lead time from 1045 to 885 minutes (by 15,31%), and waiting time – from 135 to 85 minutes (by 37,04%).

Thus the authors ascertained that despite the well-organised work-flow there still was a potential for improvement that was made obvious by application of the value stream mapping, a LEAN tool.

One of the main aims of the IT Company was to improve project success rates and overall work efficiency. The enterprise manager considered possibility to detect and reduce services which bring the least profit. It was decided to use Pareto analysis which is widely known as a useful tool to detect the most important changes to make and which is also one of the recommended LEAN tools (but is used not only by LEAN). Pareto analysis and diagrams (see Fig. 6) refer to 80/20 rule assuming that, for example, 20% of activities lead to 80% of loses.

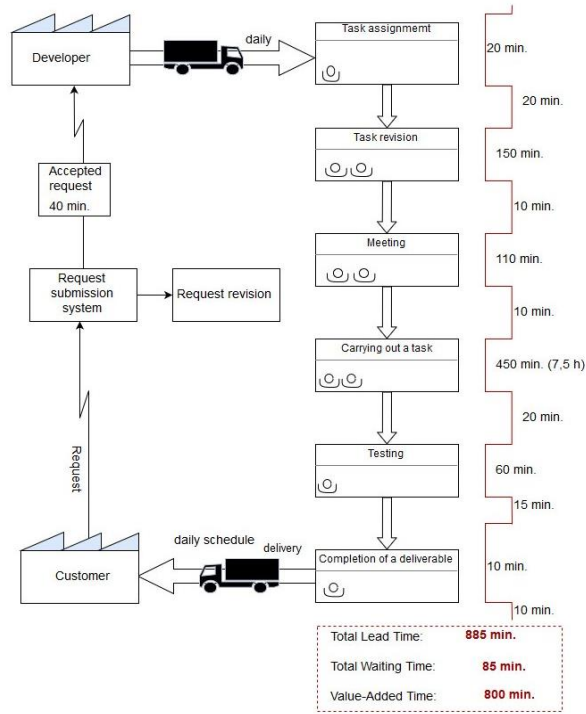

**Fig. 6**. Value Stream Mapping. Improved stream

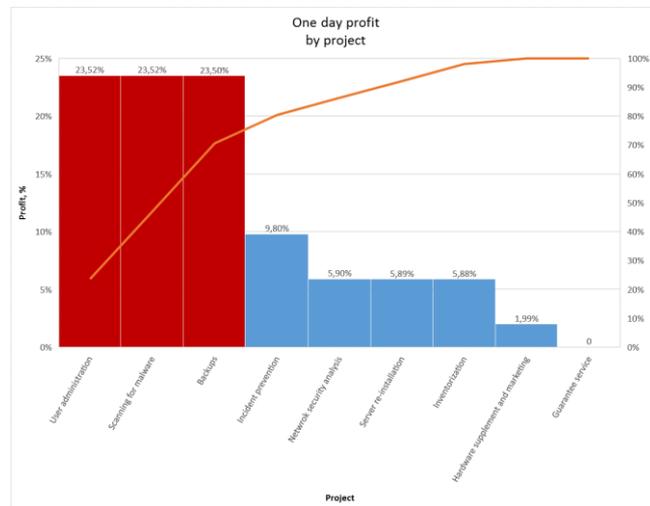

**Fig. 7**. Pareto analysis: one-day profit by project

This tool is well-known in LEAN and also in PMBOK and it is considered as a creative yet an effective tool by many experts in many areas.

The authors also analysed computer network services with an aim to identify the less or non-profitable services and consider discontinuation of these services. According to the established business processes and their terminology within the Company services are called "projects". The Pareto analysis was used to draw a diagram with the following data dimensions: each project's one-day profit as a percentage from the total, project's (service) occurrence frequency (in month) and project's (service) duration (in days). The created diagram (see Fig. 7) demonstrates which projects are more profitable than others, for example, user administration, malware detection and backuping represent only 20% of all work time of all computer network administration projects but these projects make ~80% of the total profit. While the rest 80% of the projects such as guarantee service, hardware maintenance, inventory update, server reinstallation and network security update show different profitability rate, however together they represent only ~20% of total profit. The Pareto analysis provided two main benefits, first, it enabled distinct comparison of projects and, second, the created diagram presented explicit visualisation of the results.

Combining the results of Pareto analysis with the results of value stream mapping, the authors were able to conclude that hardware maintenance and sales for this specific enterprise has the highest waiting time – 47% of whole time of realizing which is difficult to reduce and at holidays (at least two times a year) there are significantly more orders than usual which causes inhomogeneous workflow, distribution of tasks, supplement chain which cause appearing of more significantly observed wastes and negatively impacts not only the speed of specific business process or all project realization but also employee load. To sum up, there can be observed not only 'muda' and especially waiting (see Section 4), but also 'mura' - inhomogeneous workflow and even 'muri' – inhomogeneous distribution of tasks. All kinds of wastes which according to Lean must be avoided, are caused by one business process which is the most unprofitable to the Company therefore the Company might consider discontinuing this project (service) and release resources to find "smarter" usage, paying more attention and resources to such projects as scanning for malware, user management and data backups which are the most profitable to this specific Company.

Thus, the techniques, tools and analysis described in this section are useful and applicable also even in cases when organisations record high project success rates, i.e. there is always room for improvement.

## Conclusions

Though many organizations perform quite successfully and produce high quality products, project failures occur occasionally. The LEAN approach can be used very successfully to tackle the failures also in information technology companies in order to optimize results of IT projects such as computer network administration, user administration, guarantee service, etc.

The LEAN approach provides several benefits. One of them is the LEAN philosophy which penetrates all organizational processes and areas including organization's culture. Unfortunately changes in organization's culture are hard to measure, especially in short-term. Other benefit of LEAN techniques and tools is identifying wastes and potential areas for improvement even in organisations that record high performance results. Moreover, LEAN tools also provide excellent visualisation aid to identify problem areas and non-value adding activities, capture workflow, understand processes, minimize waiting time and define goals for improvement.

Value stream mapping used for improvement of Company's processes led to eliminating of wastes; the LEAN tools could be easily applied without additional resources, so, business processes improvements could be easily succeeded.

Although LEAN principles usually aren't applied on IT companies as LEAN is not traditionally taken into account as IT project management methodology, the research showed that LEAN could be applied from the fundamental steps to specific tools which satisfies specific aims.

The case study also showed that companies can already work according to specific LEAN principles or use LEAN tools or techniques without any specific knowledge about LEAN. It is easier to implement LEAN in IT companies if the first steps are already taken, for instance, if companies have already functioning quality management system. A quality management system which is certified according to the ISO 9001 standards not only doesn't contradict with LEAN but they also supplement each other (as it was in the case of the analysed Company).

Pareto analysis provides an effective tool to resolve a lot of issues identifying problem areas or non-profitable processes or activities or even projects. Another advantage of this tool is its wide applicability in many business areas to explore different reasons for issues. As it was shown in Section 5, Pareto analysis helps easily detect the most ineffective and unprofitable processes which takes the most part of resources and brings the least profit. Therefore the Pareto analysis helps choosing the most important changes to make.

Value Stream Mapping easily identified potential for reducing the total lead time from 1045 to 885 minutes (by 15,31%), and waiting time – from 135 to 85 minutes (by 37,04%).

Analysis of such a successive IT company and applying LEAN principles and techniques to it was challenging as it was hard to imagine that there could be wastes and possible improvements for the company which projects success rate is 97%. But it let to show that LEAN $5^{th}$ principle - pursue perfection according to which it is important to continuously improve results even in case of success - was working as wastes and room for improvements were detected even in the company whose projects success rate is high. All companies may have their "weak" points which are easily detected applying LEAN principles. It also is powerful in detecting and preparing list of possible improvements to eliminate detected wastes and using LEAN tools and techniques.

The provided research also shows that there are no project type limitations for LEAN as it could be applied to company with wide range of services - not only developing of software but also network administration, maintenance, sales hardware. The LEAN tools and techniques are universal for IT enterprises.

In order to detect wastes and create a realistic view of the existing situation it is important to make very thorough analysis of organization's workflow by drawing business process models and gathering statistical data. Following the research, a list of existing LEAN features and wastes should be created, evaluated and prioritized thus achieving better understanding which wastes are really harmful or have little impact on business and which must be reduced immediately.

## Acknowledgments

This work has been supported by University of Latvia project AAP2016/B032 "Innovative information technologies".